\documentclass{article}

\usepackage{microtype}
\usepackage{graphicx}
\usepackage{booktabs} 

\usepackage{hyperref}
\usepackage{subcaption}

\usepackage[accepted]{icml2023}

\usepackage{amsmath}
\usepackage{amssymb}
\usepackage{mathtools}
\usepackage{amsthm}

\usepackage[capitalize,noabbrev]{cleveref}

\theoremstyle{plain}

\theoremstyle{definition}

\theoremstyle{remark}

\usepackage[textsize=tiny]{todonotes}

\icmltitlerunning{DataCI: A Platform for Data-Centric AI on Streaming Data}

\begin{document}

\twocolumn[
\icmltitle{DataCI: A Platform for Data-Centric AI on Streaming Data}



\icmlsetsymbol{equal}{*}

\begin{icmlauthorlist}
\icmlauthor{Huaizheng Zhang}{equal,iii}
\icmlauthor{Yizheng Huang}{equal,iii}
\icmlauthor{Yuanming Li}{equal,iii}
\end{icmlauthorlist}

\icmlaffiliation{iii}{Independent Researcher}

\icmlcorrespondingauthor{Huaizheng Zhang}{zhanghuaizheng0806@gmail.com}

\icmlkeywords{Machine Learning, ICML}

\vskip 0.3in
]



\printAffiliationsAndNotice{\icmlEqualContribution} 

\begin{abstract}
We introduce DataCI, a comprehensive open-source platform designed specifically for data-centric AI in dynamic streaming data settings. DataCI provides 1) an infrastructure with rich APIs for seamless streaming dataset management, data-centric pipeline development and evaluation on streaming scenarios, 2) an carefully designed versioning control function to track the pipeline lineage, and 3) an intuitive graphical interface for a better interactive user experience. Preliminary studies and demonstrations attest to the easy-to-use and effectiveness of DataCI, highlighting its potential to revolutionize the practice of data-centric AI in streaming data contexts.
\end{abstract}

\section{Introduction}
\label{sec:introduction}

Data-centric AI has shown promising results in many domains \cite{kirillov2023segment, xu2021dataclue, liu-etal-2020-data-centric, bartolo2021models}. Compared to model-centric AI \cite{zhang2020mlmodelci, li2021modelps}, it pays more attention to data processing like data cleaning and labeling instead of model structure design, avoiding \textit{garbage in, garbage out}. Both industry and academics have responded to the growing interest in data-centric AI and made significant progress \cite{aguilar2021ease, zha2023data, mazumder2022dataperf, parrish2023adversarial}. 


However, current data-centric frameworks mainly focusing on static batch settings, are inadequate to address the unique challenges that unbounded and continuous streaming data present. First, adopting these frameworks to a streaming data environment requires much more tedious efforts, including streaming data pre-processing, data-centric function version control, etc \cite{huang2021modelcie, huang2022active}. Second, several issues unique to streaming data, such as determining the optimal frequency for running data-centric pipelines, are underexplored \cite{shankar2022rethinking, bother2023towards}.

In light of these challenges, we introduce an early-stage work, that demonstrates the potential of data-centric AI in streaming data settings, providing both practical showcases and quantitative analysis. Our contributions are two-fold:

\textbf{First}, We introduce DataCI (data continuous integration), an open-source platform that serves as a comprehensive development tool and testbed. DataCI offers a range of modular features, including a Streaming Data Sink, a Pipeline Registry, a Data-centric Function Zoo, a Pipeline Orchestration Module, and a Leaderboard. Furthermore, a graphical user interface allows users to easily manage their pipeline and visualize results, facilitating benchmark studies and the exploration of new data-centric methods for streaming data. \textbf{Second}, we provide demonstrations of DataCI and preliminary studies to highlight its user-friendly design and overall necessity.

This vision paper is an invitation to further study and enhance the application of data-centric AI to streaming data. We look forward to feedback and collaboration.

\section{System Design}

\begin{figure}[]
\vskip -0.05in
\begin{center}
\centerline{\includegraphics[width=\columnwidth]{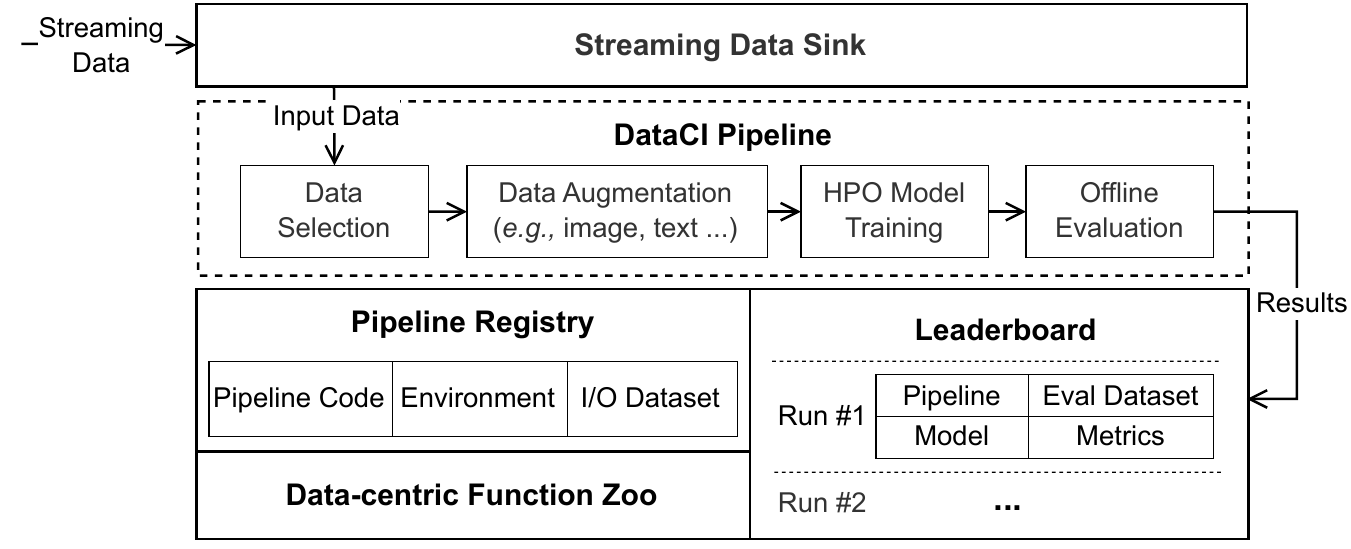}}
\caption{DataCI architecture.}
\label{fig:sys_workflow}
\end{center}
\vskip -0.4in
\end{figure}

DataCI offers an end-to-end data-centric ML experience on dynamic streaming data. As shown in Figure \ref{fig:sys_workflow}, it introduces five primary components.

\textbf{Streaming Data Sink.} DataCI uses the sink to periodically collect the data from the connected streaming system and save it into different partitions. These data partitions will be tagged according to their arrival time and used for later training and evaluation.

\textbf{Pipeline Registry.} This module enables users to build a data-centric pipeline with ease by introducing flexible APIs. These APIs cover dataset I/O, pipeline staging, and running environment management, speeding up the development of a DataCI pipeline. Furthermore, a \textbf{versioning control} function is introduced to track the pipeline lineage.

\textbf{Data-centric Function Zoo}. Different from the model hubs from HuggingFace \cite{wolf-etal-2020-transformers} and PyTorch \cite{paszke2017automatic}, we provide this module to store data processing methods such as data selection, data augmentation, and tricks applied in a specific scenario (e.g., prompting). Users can share and reuse the functions for pipeline building.

\textbf{Pipeline Orchestration.} To run a data-centric pipeline, we leverage the idea from pipeline orchestration systems like Airflow. The data flow shown in Figure \ref{fig:exp_timeline} will pass through every pre-defined stage in the pipeline and then trigger the evaluation to test the whole pipeline and know the benefit of the new data-centric function. If the pipeline further passes the A/B test, the new function or the new pipeline can be deployed to the product.

\textbf{Leaderboard.} Once the evaluation is finished, the results will be sent to the leaderboard. We use the \{run \#No.\} as the index and store the pipeline name with version, the evaluation dataset, the model name with training hype parameters, and the metric for easy reproduction and comparison.

\section{Demonstration}
\begin{figure}[]
\begin{center}
\centerline{\includegraphics[width=0.75\columnwidth]{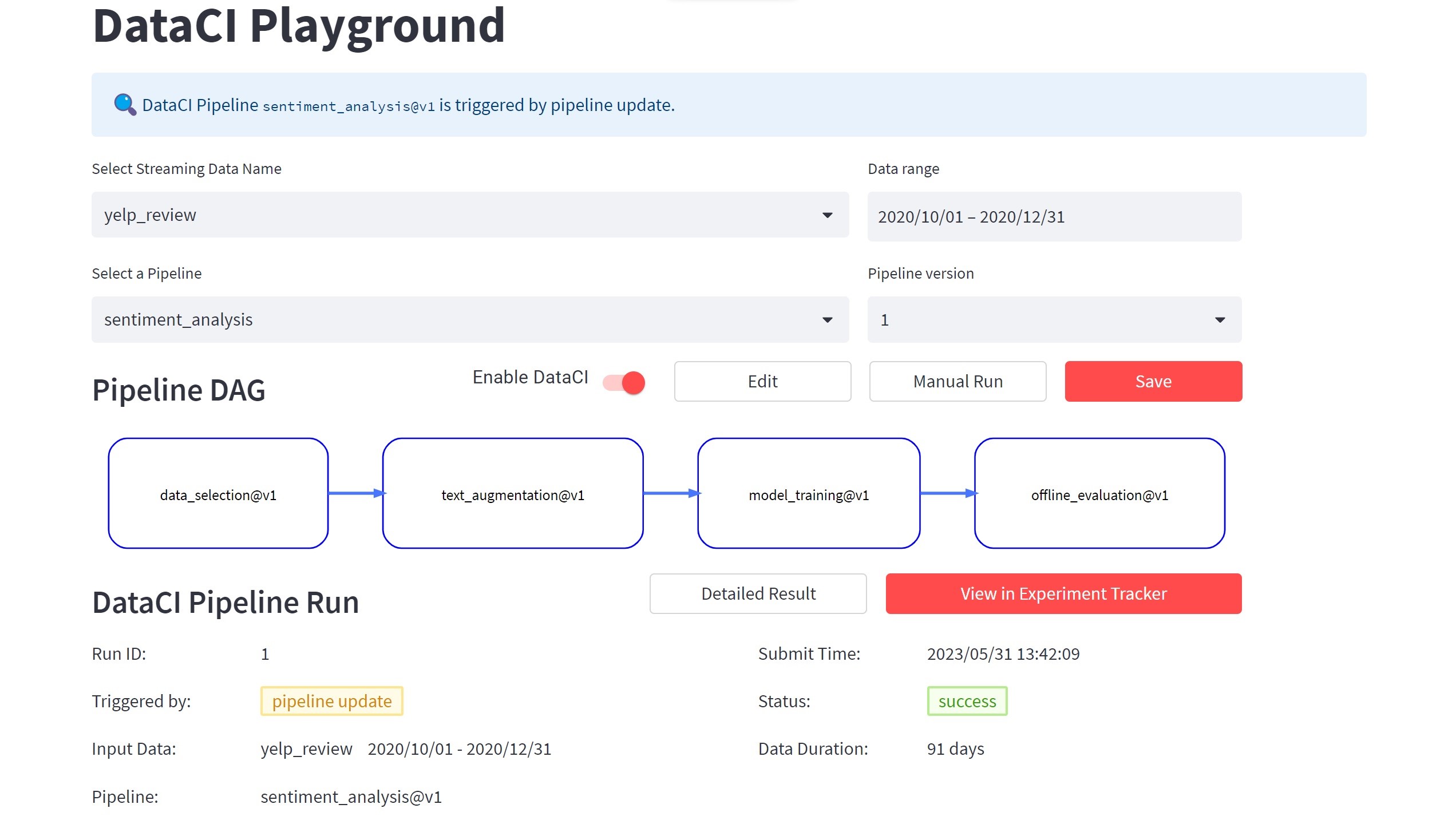}}
\caption{An interactive web interface to build, launch, compare, and visualize DataCI pipelines.}
\label{fig:dataci-playground}
\end{center}
\vskip -0.4in
\end{figure}

We demonstrate our system from two views, namely user experience investigation and quantity analysis.

\textbf{User experience.} Users' experience is our top priority. To satisfy them, we equip DataCI infrastructure with a \textbf{playground} as shown in Figure \ref{fig:dataci-playground}. This playground enables users can try our system in an interactive manner. The playground can be grouped into three sections. First, users select the data from Streaming Data Sink and pre-defined pipeline with a specific version in Pipeline Registry. Second, users can manually launch the pipeline and our playground will show a DAG (directed acyclic graph) for better visualization. Users can replace one function in the DAG and generate a new pipeline version for a quick control experiment. Third, the experiment running details will be presented in the playground for reference.

\begin{figure}[]
\begin{center}
\centerline{\includegraphics[width=\columnwidth]{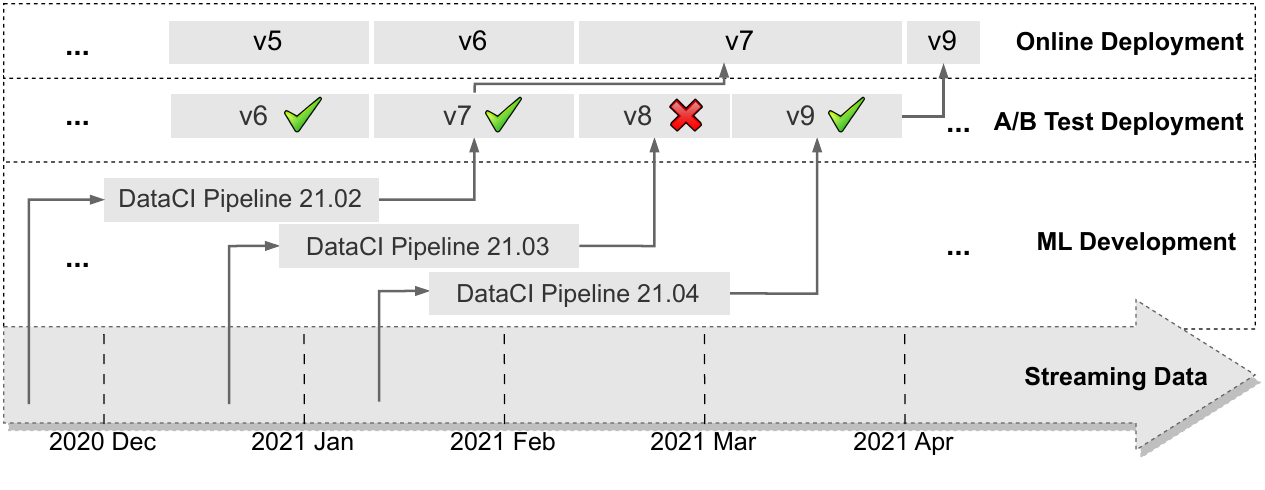}}
\caption{We use Yelp data to simulate a streaming data scenario for the DataCI evaluation.}
\label{fig:exp_timeline}
\end{center}
\vskip -0.4in
\end{figure}

\begin{figure}[b]
\vskip -0.15in
\begin{center}
\centerline{\includegraphics[width=0.75\columnwidth]{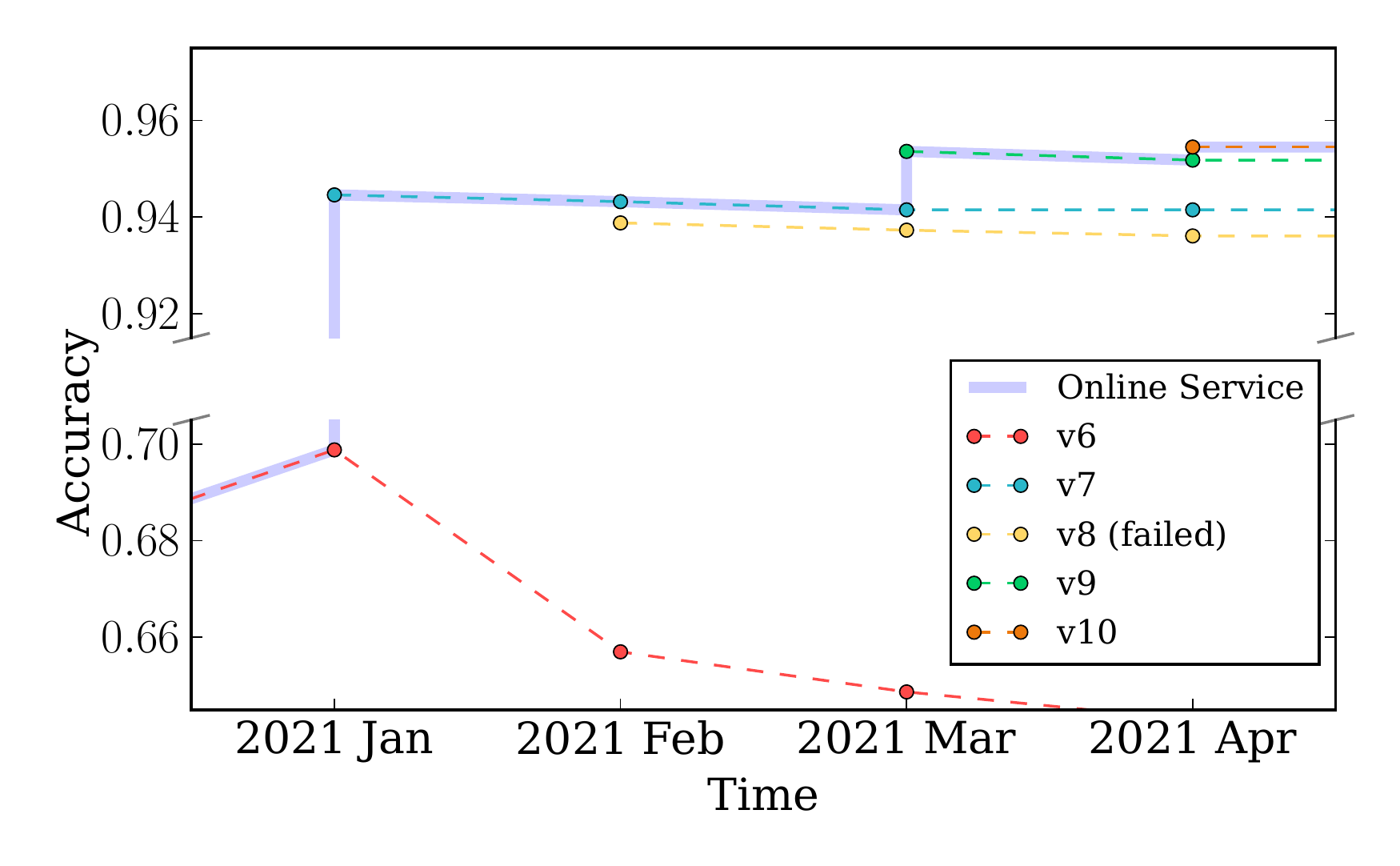}}
\caption{Performance comparison between multiple data-centric pipeline versions.}
\label{fig:exp_online_service_quality}
\end{center}
\vskip -0.2in
\end{figure}

\textbf{Quantitative analysis.} We simulate a real-world case by using Yelp dataset and sending them into our system in a streaming mode as shown in Figure \ref{fig:exp_timeline}. Assume that the online pipeline has upgraded to v5, which is our starting point. We build a pipeline v6, which passed the A/B test and was deployed to the online production. We keep using the latest data from Streaming Data Sink to develop new data-centric pipelines and test them. Only v8 fails as it can not outperform v7 as shown in Figure \ref{fig:exp_online_service_quality}. Also, from Figure \ref{fig:exp_online_service_quality}, we find if we keep using v6 without a quick pipeline update, the online performance will drop dramatically. 

This is a very preliminary study to show that a system for quick building and evaluating a data-centric pipeline on streaming data is necessary, as in the real world, data distributions change very frequently. However, this study also poses many questions that need further exploration. For example, how can we decide the upgrade frequency? Is there any better metric to measure the pipeline’s performance in this streaming scenario? etc.

\section{Conclusion}

Data-centric AI exploration has highlighted the shortcomings of existing tools in streaming data environments. To combat this, we introduced DataCI, an open-source platform that bridges these gaps. With its modular features and intuitive interface, DataCI streamlines streaming data management and method deployment. Preliminary studies affirm DataCI's potential to revolutionize data-centric AI in dynamic contexts.

\nocite{langley00}

\bibliography{example_paper}
\bibliographystyle{icml2023}



\end{document}